\documentclass[runningheads]{svmult}
\usepackage{makeidx}
\usepackage{graphicx}
\usepackage{subeqnar}
\usepackage{multicol}
\usepackage{physprbb}
\usepackage{epsfig}
\makeindex

\begin{document}%
\title*{Ferromagnetism in the Hubbard Model}
\toctitle{Ferromagnetism in the Hubbard Model}
\titlerunning{Ferromagnetism in the Hubbard Model}
\author{W.Nolting
\and M.Potthoff
\and T.Herrmann
\and T.Wegner}
\authorrunning{W.Nolting et al.}
\institute{Lehrstuhl Festk\"{o}rpertheorie, Institut f\"{u}r Physik, Humboldt-Universit\"{a}t zu Berlin}
\maketitle
\begin{abstract}
\index{abstract}We investigate the possibility and stability of bandferromagnetism in the single-band Hubbard model. This model poses a highly non-trivial
many-body problem the general solution of which has not been found up to
now. Approximations are still unavoidable. Starting from a simple
two-pole ansatz for the spectral density our approach is systematically
improved by focusing on the influence of quasiparticle damping and the
correct weak-and strong coupling behaviour. The compatibility of the
different aproximative steps with decisive moment sum rules is analysed
and the importance of a spin-dependent band shift mediated by higher
correlation functions is worked out. Results are presented in terms of
temperature- and band occupation-dependent quasiparticle densities of
states and band structures as well as spontaneous magnetisations,
susceptibilities and Curie temperatures for varying electron densities
and coupling strengths. Comparison is made to numerically essentially
exact Quantum Monte Carlo calculations recently done by other authors
using dynamical mean field theory for infinite-dimensional lattices. The
main conclusion will be that the Hubbard model provides a qualitatively
correct description of bandferromagnetism if quasiparticle damping and
selfconsistent spin-dependent bandshifts are properly taken into account.
\end{abstract}
\section{Introduction}
Ferromagnetism is bound to the existence of permanent magnetic
moments. If these belong to itinerant electrons in a partially filled
conduction band one speaks of bandferromagnetism \cite{Cap87}. Archetypical
representatives are the classical 3d ferromagnets Fe, Co, Ni. The 
microscopic  interpretation of bandferromagnetism is one of the most interesting and most complicated many-particle problems in solid state
physics. The simple but fairly successful ``Stoner criterion''
\begin{equation}
U\varrho_{0}(E_{F})>1
\end{equation}
(U: intraatomic Coulomb interaction, $\varrho_{0}$: free Bloch-density of
states (B-DOS), $E_{F}$: Fermi energy) defines as a minimum set of
ingredients for a theoretical model to describe bandferromagnetism the
Pauli-principle, the kinetic energy, the Coulomb interaction (strong and
strongly screened), and the lattice structure. 
This minimum set is realized in the Hubbard-Hamiltonian (2) for
correlated electrons on a lattice in a non-degenerate energy band,
\begin{equation}
\hat{H}=\sum\limits_{ij,\sigma}(T_{ij}-\mu\delta_{ij})c^+_{i\sigma}c_{j\sigma}+
\frac{1}{2}U\sum_{i\sigma}n_{i\sigma}n_{i-\sigma}
\end{equation}
($\mu$: chemical potential). The Coulomb interaction is restricted to 
its intraatomic part only
$(n_{i\sigma}=c^+_{i\sigma}c_{i\sigma})$, while the kinetic energy
contains the hopping integrals $T_{ij}$ being strongly related to the
lattice structure. The Pauli principle is guaranteed by the formalism of
second quantization. $c^+_{i\sigma}(c_{i\sigma})$ is the creation
(annihilation) operator for an electron with spin $\uparrow$ or
$\downarrow$ at lattice site $\vec{R}_i$.

The physics of the Hubbard-model is
decisively determined by the Coulomb coupling $U/W$ (W:
Bloch-bandwidth), the lattice structure and the band occupation
$n=\frac{1}{N}\sum_\sigma<n_\sigma>$ ($0\le{n}\le{2}$; N: number of
lattice sites). 

In spite of its simple structure the Hamiltonian (2) provokes a rather
sophisticated many-body problem, that could be solved in the past only
for a few special cases. In general approximations are unavoidable. So
it was not fully clarified until recently whether or not ferromagnetism
is possible in the Hubbard-model for finite U, finite temperatures T and
band occupations away from half filling n=1. Nagaoka showed \cite{Nag66} 
for the
special case $U=\infty$ and electron numbers $N_{e}=N\pm1$ on an sc or
bcc lattice and $N_e=N+1$ on an fcc lattice, respectively, that the fully
spin-polarized particle system represents the ground state. However,
 the (numerical) proof of finite-temperature ferromagnetism in an extended
parameter region was only recently given by Ulmke \cite{Ul98} for an infinite
dimensional $(d=\infty)$ fcc-type lattice by use of Quantum Monte Carlo (QMC)
calculations in connection with Dynamical Mean Field Theory (DMFT)
\cite{MV89,M-H89}.
On the other hand, the certainty that ferromagnetism exists in the Hubbard model
does not at all mean that the phenomenon itself is understood. What is
the physical mechanism enforcing ferromagnetic spin order of itinerant
electrons? This question can be answered, if at all, better by
analytical approaches to the highly complicated electron correlation problem than by
purely numerical evaluations. Starting from some basic features we shall
construct in the following chapters three different theories for the
Hubbard model to contribute to an answer of the above question. The
theories are constructed in such a way that each differs from the
preceding  one by eliminating an obvious shortcoming. We start from a
simple spectral density approach (SDA) which suggests itself by some rigorous
strong coupling features. However, it suffers from a complete neglect of
quasiparticle damping and a wrong weak-coupling behaviour. By use of a modified alloy analogy (MAA) the advantages of the SDA are retained but
quasiparticle damping is included. Comparison of the SDA- and the
MAA-results helps us to recognize the influence of quasiparticle damping
on magnetic stability. The weak coupling behaviour, however, remains
insufficient. From a modified perturbation theory (MPT) which correctly
reproduces the weak- as well as the strong-coupling limit, we learn by
comparison with SDA and MAA how decisive the correct weak coupling (low
energy) description is for a theoretical approach that aims at the
strong-coupling phenomenon ferromagnetism.
\section{Hubbard model basic features}
It is commonly accepted that ferromagnetism is a strong-coupling
phenomenon. A theoretical approach should therefore be reliable first of
all in the ${U\gg W}$ regime. Practically all information we are  interested
in can be derived from the retarded single-electron Green's function
\cite{Mah90}
\begin{equation}
G_{\vec{k}\sigma}(E)={\langle\langle {c_{\vec{k}\sigma};c^+_{{\vec{k}\sigma}}}\rangle\rangle}_E=-\I \int\limits_{0}^{\infty}dt\exp({\frac{\I}{\hbar}Et})\langle{[c_{\vec{k}\sigma}(t),c^+_{\vec{k}\sigma}(0)]_+}\rangle
\end{equation}
\begin{equation}
c_{\vec{k}\sigma}={\frac{1}{\sqrt{N}}\sum_{i}c_{i\sigma}\exp(-\I {\vec{k}\vec{R}_i}})
\end{equation}
$[..,..]_{+(-)}$ means the anticommutator (commutator) and $\langle .. \rangle$
the grandcanonical average. $\vec{k}$ is a wavevector from the first
Brillouin zone. Of equivalent importance is the single-electron spectral
density being directly related to the bare line shape of an angle- and
spin-resolved (direct or inverse) photoemission experiment:
\begin{equation}
S_{\vec{k}\sigma}(E)=-{\frac{1}{\pi}}ImG_{\vec{k}\sigma}(E+\I 0^+)
\end{equation}
According to the pioneering work of Harris and Lange \cite{HL67} we know that
in the strong coupling limit $S_{\vec{k}\sigma}(E)$ is built up by two
main peaks near $T_0$ and $T_0+U$ where $T_0$ is the centre of gravity
of the free Bloch band, and additional satellite peaks at $T_0+pU$
$(p=-1,-2,\ldots ; p=2,3,\ldots )$ (s. Fig.1). The spectral weights ($\propto$ peak area) of the
satellites, however, decrease rapidly with increasing distance from the
main peaks. Already the next neighbours $(p=-1,+2)$ acquire only a weight
of order $({\frac{W}{U}})^4$ being therefore negligible in the
strong-coupling regime. So we can  assume for ${U\gg W}$ a two-peak structure of
the spectral density.
The exact shapes of the peaks are not known, but the
centres of gravity are, 
\begin{equation}
T_{1\sigma}(\vec{k})=T_0+(1-\langle n_{-\sigma}\rangle)
(\varepsilon(\vec{k})-T_0)+\langle n_{-\sigma}\rangle B_{\vec{k}-\sigma}+0({\frac{W}{U}})^4
\end{equation}
\begin{equation}
T_{2\sigma}(\vec{k})=T_{0}+U+\langle n_{-\sigma}\rangle
(\varepsilon(\vec{k})-T_0)+(1-\langle n_{-\sigma}\rangle )B_{\vec{k}-\sigma}+0(\frac{W}{U})^4
\end{equation}
as well as their spectral weights:
\begin{equation}
\alpha_{1\sigma}(\vec{k})=(1-\langle n_{-\sigma}\rangle )+0(\frac{W}{U})=1-\alpha_{2\sigma}(\vec{k})
\end{equation}
\begin{figure}[t]
  \begin{center}
    \epsfig{file=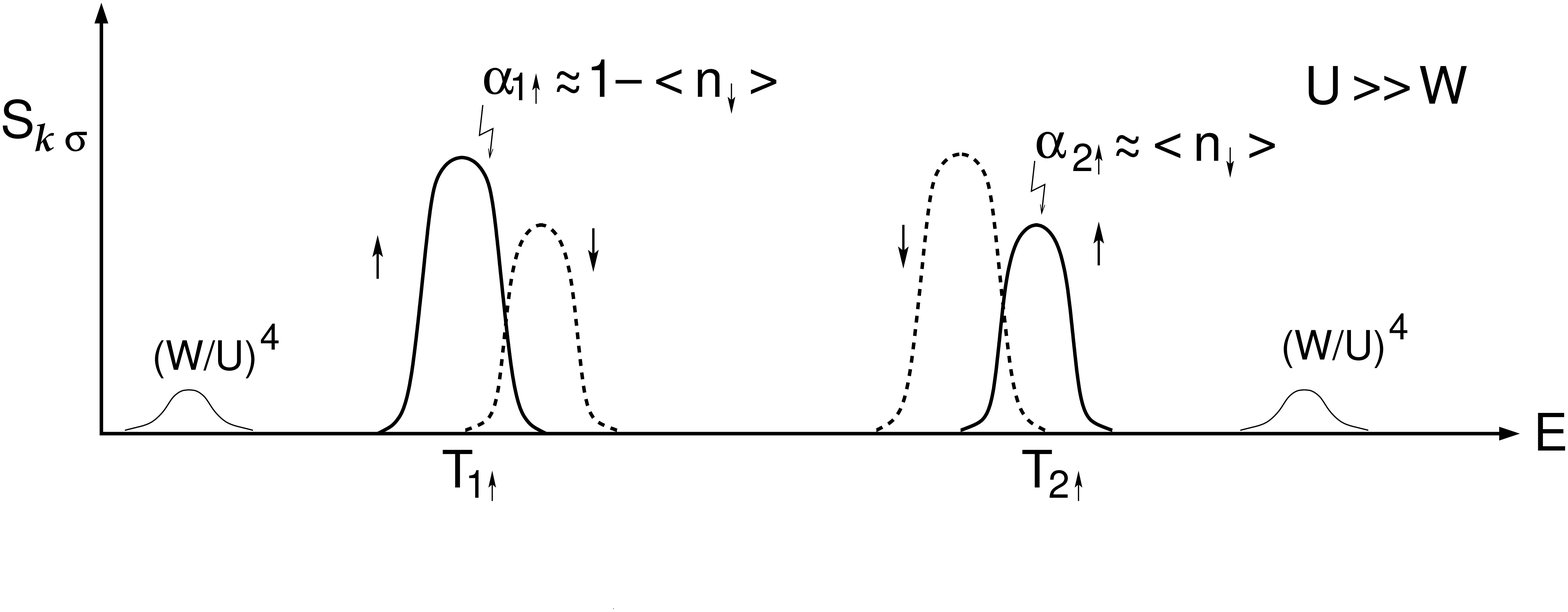,scale=0.25}
    \caption{Single electron spectral density of the Hubbard model in 
      the strong coupling regime ($U\gg W$) as function of energy.}
    \label{fig:Nol:harlang}
  \end{center}
\end{figure}
At least formally the centres of gravity carry a spin-dependence which
may give rise to an additional exchange splitting of the two main peaks
as the fundamental precondition for ferromagnetism. In the following it
is demonstrated that such a spin asymmetry is mainly due to the ``band
correction'' $B_{\vec{k}-\sigma}$:
\begin{equation}
B_{\vec{k}-\sigma}=B_{-\sigma}+F_{\vec{k}-\sigma}
\end{equation}
The local term $B_{-\sigma}$ (``band-shift'') can be interpreted as a
correlated electron hopping:
\begin{equation}
\langle n_{-\sigma}\rangle (1-\langle n_{-\sigma}\rangle)B_{-\sigma}={\frac{1}{N}}\sum\limits_{i,j}^{i{\neq
    j}}T_{ij}\langle c_{i-\sigma}^+c_{j-\sigma}(2n_{i\sigma}-1)\rangle
\end{equation}
For $U\gg W$ and less than half-filled bands double occupancies are very
unlikely so that the second term in the bracket dominates. That means
that the shift of the $\sigma$-spectrum is correlated with the negative
kinetic energy of the $(-\sigma)$-electrons. For further evaluations it
will turn out to be decisive that the band-shift $B_{-\sigma}$ can
exactly be expressed by the single-electron spectral-density \cite{GN88,NB89}
\begin{eqnarray}
  \lefteqn{\langle n_{-\sigma}\rangle (1-\langle n_{-\sigma}\rangle
    )B_{-\sigma}= }\nonumber\\&& {\frac{1}{N\hbar}}\sum_{\vec{k}}
  (\varepsilon(\vec{k})-T_0)\int^{\infty}_{-\infty}
  \D Ef_{-}(E)({\frac{2}{U}}(E-\varepsilon(\vec{k}))-1))
  S_{\vec{k}-\sigma}(E-\mu)
\end{eqnarray}
$f_{-}(E)=(\exp{\beta (E-\mu)}+1)^{-1}$ is the Fermi function. $B_{-\sigma}$
obviously disappears in the zero-bandwidth
limit$(\varepsilon(\vec{k})\longrightarrow T_0\, \forall\,\vec{k})$.

The k-dependent part of the band correction (9) is built up by
density-density, double hopping and spinflip correlation terms
\begin{eqnarray}
\lefteqn{\langle n_{-\sigma}\rangle (1-\langle n_{-\sigma}\rangle )F_{\vec{k}-\sigma}=
  }\nonumber\\&&{\frac{1}{N}}\sum\limits_{i,j}^{i\neq j}T_{ij}\exp(-\I
\vec{k}(\vec{R}_i-\vec{R}_j))(\langle n_{i-\sigma}n_{j-\sigma}\rangle
-\langle n_{j-\sigma}\rangle ^2\nonumber\\&& +\langle c^+_{j-\sigma}
c^+_{j\sigma}c_{i-\sigma}c_{i\sigma}\rangle +\langle c^+_{j\sigma}c_{j-\sigma}c^+_{i-\sigma}c_{i\sigma}\rangle)
\end{eqnarray}
It vanishes in the zero-bandwidth limit and has no direct influence on
the centres of gravity $T_{1,2\sigma}$ of the quasiparticle subbands
(Hubbard bands) that emerge from the two spectral density peaks (Fig. 1):
\begin{equation}
\hat T_{1\sigma}=
{\frac{1}{N}}\sum_{\vec{k}} T_{1\sigma}(\vec{k})=T_0+\langle
n_{-\sigma}\rangle B_{-\sigma}
\end{equation}
\begin{equation}
\hat T_{2\sigma}={\frac{1}{N}}\sum_{\vec{k}}
T_{2\sigma}(\vec{k})=T_0+U+(1-\langle n_{-\sigma}\rangle )B_{-\sigma}
\end{equation}
$F_{\vec{k}-\sigma}$ may, however, lead to a spin-dependent bandwidth
correction competing then with the other band narrowing terms in (6) and
(7), respectively. The importance of $F_{\vec{k}-\sigma}$ with respect to
ferromagnetism shall be discussed in Sect.III. Unfortunately, it cannot be
expressed by the spectral density as the band shift $B_{-\sigma}$ by
eq. (11). A determination of $F_{\vec{k}-\sigma}$ therefore requires further
approximations. 

If $B_{\vec{k}-\sigma}$ has indeed such vital implications a theoretical
approach should handle this term with special care. That can effectively
be controlled by the spectral moments of the spectral density,
\begin{eqnarray}
M^{(n)}_{\vec{k}\sigma}={\frac{1}{\hbar}}\int\limits_{-\infty}^{\infty}\D
EE^nS_{\vec{k}\sigma}(E)\quad n=0,1,2,3,\ldots
\end{eqnarray}
which can be calculated independently of $S_{\vec{k}\sigma}(E)$ via
\begin{equation}
M^{(n)}_{\vec{k}\sigma}=\langle [(\I \hbar \frac{\partial}{\partial
  t})^nc_{\vec{k}\sigma}(t),c^+_{\vec{k}\sigma}(t')]_+\rangle _{t=t'}
\end{equation}
In practice, however, the moments are useful only for low order n
because with increasing n eq.(16) produces higher expectation values
which are usually unknown and not expressible by the spectral density,
either. The important band correction $B_{\vec{k}-\sigma}(E)$ first
appears in $M^{(3)}_{\vec{k}\sigma}$ \cite{GN88,NB89}.
\begin{eqnarray}
  M^{(0)}_{\vec{k}\sigma}=1\\
  M^{(1)}_{\vec{k}\sigma}=\hat{\varepsilon}(\vec{k})+U\langle
  n_{-\sigma}\rangle \\
  M^{(2)}_{\vec{k}\sigma}=\hat{\varepsilon}^2
  (\vec{k})+
  2{\hat{\varepsilon}}(\vec{k})U\langle n_{-\sigma}\rangle +U^2
  \langle n_{-\sigma}\rangle \\
  \lefteqn{M^{(3)}_{\vec{k}\sigma}=
    \hat{\varepsilon}^3
  (\vec{k})+3\hat{\varepsilon}^2(\vec{k})U\langle n_{-\sigma}\rangle
  +\hat{\varepsilon}(\vec{k})U^2\langle n_{-\sigma}\rangle (2+\langle
  n_{-\sigma}\rangle )}\\\quad\quad +U^3\langle n_{-\sigma}\rangle
+U^2\langle n_{-\sigma}\rangle (1-\langle n_{-\sigma}\rangle )
(B_{\vec{k}-\sigma}+T_0-\mu)\nonumber
\end{eqnarray}
($\hat{\varepsilon}(\vec{k})=\varepsilon(\vec{k})-\mu$). A necessary 
condition for a theoretical approach to be consistent with
the strong-coupling results (6),(7),(8) is that the first four moments
$M^{(n)}_{\vec{k}\sigma}, n=0,1,2,3$ are correctly reproduced. The
condition becomes even sufficient when additionally the
zero-bandwidth limit \cite{Hub63} is fullfilled. One can elegantly check 
 the strong coupling consistency by use of the high-energy expansion 
of the single-electron Green's function
\begin{equation}
  G_{\vec{k}\sigma}(E)=\int\limits_{-\infty}^{+\infty}
  dE'\frac{S_{\vec{k}\sigma}(E')}{E-E'}\longrightarrow
  \hbar\sum\limits_{n=0}^{\infty}\frac{M^{(n)}_{\vec{k}\sigma}}{E^{n+1}}
\end{equation}
which can also be used in the Dyson equation,
\begin{equation}
  G_{\vec{k}\sigma}(E)=G^{(0)}_{\vec{k}}(E)+
  G^{(0)}_{\vec{k}}(E){\frac{1}{\hbar}}\Sigma_{\vec{k}\sigma}(E)
  G_{\vec{k}\sigma}(E),
\end{equation}
where
\begin{equation}
G^{(0)}_{\vec{k}}(E)=\hbar(E+\mu-\varepsilon(\vec{k}))^{-1}
\end{equation}
is the $U=0$\, Green's function,
to yield a respective expansion for the selfenergy:
\begin{equation}
\Sigma_{\vec{k}\sigma}(E)=\sum\limits_{m=0}^{\infty}{\frac{C^{(m)}_{\vec{k}\sigma}}{E^m}}
\end{equation}
The coefficients
$C^{(m)}_{\vec{k}\sigma}=f_{m}(M^{(0)}_{\vec{k}\sigma},\ldots,M^{(m+1)}_{\vec{k}\sigma})$
are simple functions of the moments up to order $m+1$:
\begin{eqnarray}
C^{(0)}_{\vec{k}\sigma}=M^{(1)}_{\vec{k}\sigma}-(\varepsilon(\vec{k})-\mu)\\
C^{(1)}_{\vec{k}\sigma}=M^{(2)}_{\vec{k}\sigma}-(M^{(1)}_{\vec{k}\sigma})^2\\
C^{(2)}_{\vec{k}\sigma}=M^{(3)}_{\vec{k}\sigma}-2M^{(2)}_{\vec{k}\sigma}M^{(1)}_{\vec{k}\sigma}+(M^{(1)}_{\vec{k}\sigma})^3\\
\ldots \nonumber
\end{eqnarray}
For a given theoretical approach one can easily control whether or not
the leading term in the expansions (21)and (24), respectively, match with the
rigorously calculated moments. However, even the opposite procedure may
be useful, infering approximately the overall energy dependence of the
Green's function and the selfenergy, respectively, from the first few 
exactly calculated terms. A proposal how this can be done is exemplified in the next Section.

Rigorous statements are of course also possible in the weak coupling
regime ($U\ll W$). As mentioned in Sect.I it remains a challenging
problem to find out how decisive weak coupling and low-energy properties
are for a theoretical approach to correctly describe the strong-coupling
phenomenon ``ferromagnetism''. In the ``diagram-language'' the Dyson equation (22) reads
\begin{figure}[htp]
  \begin{center}
    \includegraphics[width=10.5cm]{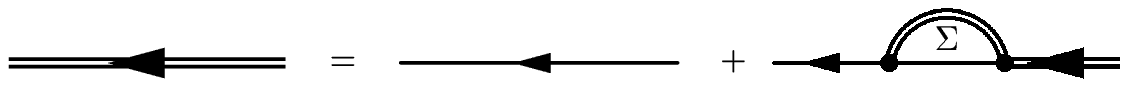}
    \label{fig:Nol:Dyson}
  \end{center}
\end{figure} \vspace{-1cm}
\begin{eqnarray}
 \qquad G_{\vec{k}\sigma}(E)\qquad\; =\qquad\quad
  G^{(0)}_{\vec{k}}(E)\qquad 
  +\quad G^{(0)}_{\vec{k}}(E){\frac{1}{\hbar}}
  \Sigma_{\vec{k}\sigma}(E)G_{\vec{k}\sigma}(E)
\nonumber
\end{eqnarray}
Using standard perturbation theory \cite{Mah90,NoltingBd7,AGD64} 
one has to sum up for the
selfenergy $\Sigma_{\vec{k}\sigma}(E)$ all dressed skeleton diagrams.
A ``skeleton diagram'' is a selfenergy diagram that does not contain any
selfenergy insertion in its propagators. If in addition the propagators
are ``full'' Green's functions $G_{\vec{k}\sigma}(E)$ then the skeleton
is ``dressed''. That means for the Hubbard model up to second order:
\vspace{-0.5cm}
\begin{figure}[h]
  \begin{center}
    \includegraphics[scale=0.75]{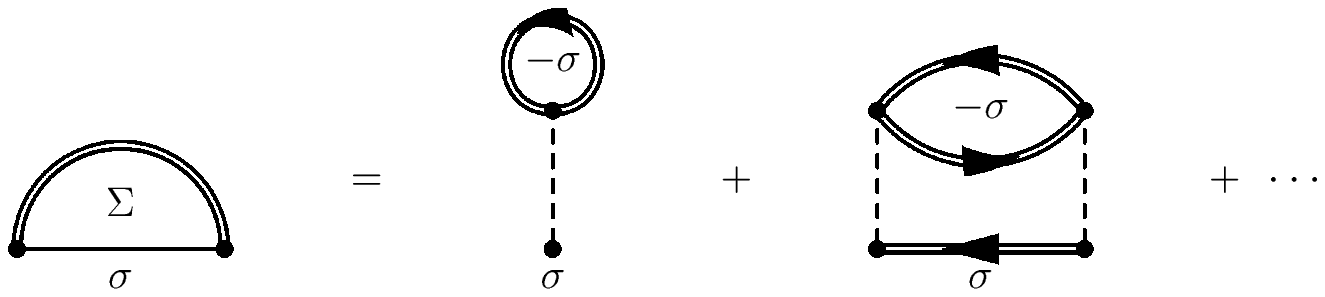}
    \label{fig:Nol:Selfenergy}
  \end{center}
\end{figure}
\newline
Conventional diagram rules \cite{Mah90,NoltingBd7,AGD64} 
yield as selfconsistent second order perturbation theory in$\frac{U}{W}$:
\begin{equation}
\Sigma_{\vec{k}\sigma}(E)=U\langle n_{-\sigma}\rangle +
U^2\Delta_{\vec{k}\sigma}(E)+0(U^3)
\end{equation}
The linear term is the Hartree-Fock (Stoner) part while the more
complicated second order contribution is in general non-local, complex, and 
energy dependent:
\begin{eqnarray}
  \lefteqn{\Delta_{\vec{k}\sigma}(E)=}\nonumber\\&&\frac{1}{\hbar^3N^2}
  \sum\limits_{\vec{q}\vec{p}}\int\int\int\D\varepsilon_1\D\varepsilon_2
  \D\varepsilon_3{\frac{S_{\vec{q}+\vec{k}\sigma}(\varepsilon_1-\mu)
      S_{\vec{p}-\sigma}(\varepsilon_2-\mu)S_{\vec{p}-\vec{q}-\sigma}
      (\varepsilon_3-\mu)}{E+\mu-\varepsilon_1+\varepsilon_2-\varepsilon_3
      +\I 0^+}}\nonumber\\&&\times \big( f_-(\varepsilon_1)f_-(-\varepsilon_2)
  f_-(\varepsilon_3)+f_-(-\varepsilon_1)f_-(\varepsilon_2)
  f_-(-\varepsilon_3)\big)
\end{eqnarray}
S is the full spectral density which has to be determined
self-consistently. Several alternatives are thinkable which are correct
and equivalent up to order $U^2$. One of these is to replace the
``full'' spectral density in (29) by its Hartree-Fock version \cite{SC90,ZH83}
\begin{equation}
S^{(1)}_{\vec{k}\sigma}(E)=S^{(0)}_{\vec{k}\sigma}(E-U\langle
n_{-\sigma}\rangle ^{(1)})
\end{equation}
To avoid an ambiguity we require that the Hartree-Fock particle
densities are the same as those in the full calculation:
\begin{equation}\langle n_{\sigma}\rangle ^{(1)}=\langle n_\sigma\rangle
\end{equation}
that can be regulated by considering the chemical potential $\mu^{(1)}$
in (30) as a proper fit parameter. There are of course other
possibilities to avoid the mentioned ambiguity \cite{KK96}. 

Several very important conclusions can be drawn from (28), e.g. for the low-energy
behaviour of the selfenergy \cite{Lut61} at $T=0\, (\alpha ,\beta
,\gamma,\,real)$,
\begin{equation}
\Sigma_{\vec{k}\sigma}(E)=\alpha_{\vec{k}\sigma}+\beta_{\vec{k}\sigma}E+\I\vec{\gamma}_{\vec{k}\sigma}E^2+\ldots 
\end{equation}
That means that the imaginary part of $\Sigma$ disappears at the Fermi
edge ($E=0$) indicating quasiparticles with infinite lifetimes. At low
but finite temperatures holds
\begin{equation}
Im\Sigma_{\vec{k}\sigma}(0)\propto T^2\quad(T\longrightarrow 0)
\end{equation}
(32) and (33) are necessary to guarantee the correct Fermi liquid behaviour.
\section{Analytical Approaches}
We can now use the basic features of the last section to construct and
control three different analytical approaches. The main goal is to work
out the essentials for ferromagnetism in the Hubbard model by a critical
comparison of these theories.
\subsection{Spin-dependent band shift}
Having in mind the two-peak structure of the spectral density in the
strong-coupling regime (Fig.1) we can construct a simple ``spectral
density approach'' (SDA). If we initially assume that quasiparticle
damping, responsible for the finite peak widths, does not play a
dominant role for magnetic properties then the following two-pole ansatz
appears to be plausible:
\begin{equation}
S_{\vec{k}\sigma}(E)=\hbar\sum\limits_{j=1}^{2}\alpha_{j\sigma}(\vec{k})\delta
(E+\mu -E_{j\sigma}(\vec{k}))
\end{equation}
The quasiparticle energies $E_{j\sigma}(\vec{k})$ and their spectral
weigths $\alpha_{j\sigma}(\vec{k})$ are easily fixed by equating the
first four moments. The resulting selfenergy has a remarkable structure
\cite{GN88,NB89,HN97,Ro69}, in particular the important ``band correction''
$B_{\vec{k}-\sigma}$ is involved:
\begin{equation}
\Sigma^{SDA}_{\vec{k}\sigma}(E)=U\langle n_{-\sigma}\rangle
\frac{E+\mu-T_0-B_{\vec{k}-\sigma}}{E+\mu-T_0-B_{\vec{k}-\sigma}-U(1-\langle
  n_{-\sigma}\rangle)}
\end{equation}
According to (10) and (12) $B_{\vec{k}-\sigma}$ vanishes in the
zero-band-width limit ($W\longrightarrow
0$). $\Sigma^{SDA}_{\vec{k}\sigma}(E)$ is then the exact $W=0$ selfenergy
\cite{Hub63,HN97}. Additionally the SDA fullfills by 
construction the first four
moments so that the strong-coupling behaviour must be correct according
to (6)-(8).The first three terms of the high-energy expansion of
$\Sigma^{SDA}_{\vec{k}\sigma}(E)$ agree exactly with (25)-(27).

We note in passing that the so-called ``Hubbard-I approach'' \cite{Hub63} leads
to a selfenergy being identical to that of the $W=0$ limit. It thus
coincides with that of the SDA for $B_{\vec{k}-\sigma}=0$. Comparing the
results of SDA and Hubbard-I therefore helps to understand the
implications due to $B_{\vec{k}-\sigma}$.

Two severe shortcomings of the SDA are also very obvious. The
first is the neglect of quasiparticle damping by the delta-function
ansatz (34) which leads to a real selfenergy
$\Sigma^{SDA}_{\vec{k}\sigma}(E)$. The second concerns the weak-coupling
regime, which is surely violated by the strong-coupling theory SDA. It is
not to expect that Fermi liquid behaviour is correctly described by the
SDA. These two points shall be attacked and eliminated by the subsequent
approaches. But let us first inspect what the SDA tells about
bandferromagnetism in the Hubbard model.

For a full SDA-solution we have to determine the ``band
correction'' $B_{\vec{k}-\sigma}$(9). The local part  $B_{-\sigma}$ does
not need a special treatment because of its direct connection (11) to
the spectral density itself. The $\vec{k}$-dependent part
$F_{\vec{k}-\sigma}$ appears a bit more complicated. Assuming
translational invariance and next neighbour hopping, only, the
$\vec{k}$-dependence can be separated \cite{HN97,Ro69}:
\begin{eqnarray}
\langle n_{-\sigma}\rangle (1-\langle n_{-\sigma}\rangle)F_{\vec{k}-\sigma}=(\varepsilon(\vec{k})-T_0)\sum\limits_{i=1}^{3}F^{(i)}_{-\sigma}\\
F^{(1)}_{-\sigma}=\langle n_{i-\sigma}n_{j-\sigma}\rangle-\langle
n_{-\sigma}\rangle ^2\\
F^{(2)}_{-\sigma}=\langle
c^+_{j-\sigma}c^+_{j\sigma}c_{i-\sigma}c_{i\sigma}\rangle \\
F^{(3)}_{-\sigma}=\langle c^+_{j\sigma}c_{j-\sigma}c^+_{i-\sigma}c_{i\sigma}\rangle
\end{eqnarray}
i,j are numbering next neighbours. The method used to determine the
higher correlations $F^{(i)}_{-\sigma}$ shall be exemplified for
$F^{(3)}_{-\sigma}$. First we rewrite $F^{(3)}_{-\sigma}$ as
\begin{equation}
F^{(3)}_{-\sigma}=\sum\limits_l\delta_{jl}\langle c^+_{l\sigma}c_{j-\sigma}c^+_{j+\Delta
  -\sigma}c_{j+\Delta\sigma}\rangle
\end{equation}
where the index $\Delta$ corresponds to the lattice vector which
connects two neighbouring sites $\vec{R}_i$ and $\vec{R}_j$. Because of
translational invariance $F^{(3)}_{-\sigma}$ does not depend on the
explicit value of $\Delta$. We now introduce a ``higher'' spectral
density $S^{(3)}_{\vec{k}\sigma}(E)$ as the ($\vec{k},E$)-dependent
Fourier transform of
\begin{equation}
S^{(3)}_{jl\sigma}(t,t')=\frac{1}{2\pi}\langle [(c_{j-\sigma}c^+_{j+\Delta
  -\sigma}c_{j+\Delta\sigma})(t),c^+_{l\sigma}(t')]_{+}\rangle
\end{equation}
The spectral theorem yields:
\begin{equation}
F^{(3)}_{-\sigma}=\frac{1}{\hbar N}\sum\limits_{\vec{k}}
\int\limits_{-\infty}^{\infty}\D
Ef_-(E)S^{(3)}_{\vec{k}\sigma}(E-\mu)
\end{equation}
According to the definition (41) the poles of
$S^{(3)}_{\vec{k}\sigma}(E)$ must belong to the single-electron
excitations of the Hubbard system. From the spectral representation of
$S^{(3)}_{\vec{k}\sigma}(E)$ and by comparison with the respective
representation of the single-electron spectral density $S_{\vec{k}\sigma}(E)$  
again a two pole ansatz appears to be consistent,
\begin{equation}
S_{\vec{k}\sigma}(E)=\hbar\sum\limits_{j=1}^{2}\hat\alpha_{j\sigma}(\vec{k})\delta
(E+\mu -E_{j\sigma}(\vec{k}))
\end{equation}
where the quasiparticle energies $E_{j\sigma}(\vec{k})$ are the same as
those 
in (34) so that the spectral weights $\hat\alpha_{j\sigma}$ are the only
unknown parameters. They are fixed by the first two spectral moments of
$S^{(3)}_{\vec{k}\sigma}(E)$ leading via (42) then to an (approximate)
expression for $F^{(3)}_{-\sigma}$.

In an analogous manner the correlation terms $F^{(1)}_{-\sigma}$ and
$F^{(2)}_{-\sigma}$ are determined by two-pole ansatzes of properly
chosen ``higher'' spectral densities $S^{(1,2)}_{\vec{k}\sigma}$
\cite{HN97}. We note in passing that the same procedure can of course also be
applied to the term $\langle
n_{i\sigma}c^+_{i-\sigma}c_{j-\sigma}\rangle$ in the ``band shift'' $B_{-\sigma}$ (10) yielding then the exact
result (11).

Fig.2 demonstrates for an sc lattice that the just-described SDA allows
for bandferromagnetism in the Hubbard model. The zeros of the inverse
paramagnetic susceptibility $\chi^{-1}$ as a function of the
band occupation n indicate instabilities of the paramagnetic state
towards ferromagnetism $(T=0K)$. The instabilities appear as soon as U
exceeds the critical value $U_C \approx 4W$. It is a special feature  of
the SDA \cite{HN97,THerN97}, maybe even of the Hubbard model itself, that there
appear for $U > U_C$ two ferromagnetic solutions, i.e. two zeros of
$\chi^{-1}$. The first solution sets in at $n_C \ge{0.34}$ 
where the actual value only slightly depends on
U running into saturation  (m=n) for $n \ge 0.68$ (see inset Fig.2). The
second solution appears for higher band occupation, but does never reach
saturation and is always less stable than the other solution.
\begin{figure}[t]
  \begin{center}
    \epsfig{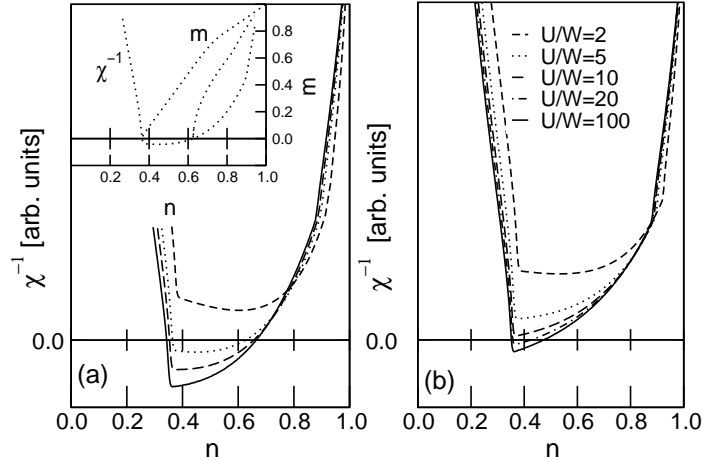}
    \caption{Inverse paramagnetic static susceptibility $\chi^{-1}$ for
      the sc lattice as a function of the band occupation n for various
      values of U. (a) System with the full $\vec{k}$-dependent
      selfenergy. (b) System with a local selfenergy
      ($F_{\vec{k}-\sigma}\equiv 0$). The inset shows the magnetic
      moment as a function of n for $U/W=5$.\vspace{-0.3cm}}
    \label{fig:No:sdachi}
  \end{center}
\end{figure}
Two important aspects of the SDA solution in Fig.2 should be
stressed. The first concerns the comparison with the corresponding
results of the Hubbard-I approach \cite{Hub63}, which does not allow
ferromagnetism on the sc lattice. On the other hand, the Hubbard-I
selfenergy differs from (35) ``only'' by neglecting the
``band correction'' $B_{\vec{k}-\sigma}$. Consequently, the stronger
magnetic stability in the SDA must be due to $B_{\vec{k}-\sigma}$. The
second remark aims at the non-locality ($\vec{k}$-dependence) of the
electronic selfenergy provoked by the ``bandwidth correction''
$F_{\vec{k}-\sigma}$ (36) in $B_{\vec{k}-\sigma}$. 
Part (b) of Fig.2 demonstrates
its importance. Switching off this term leads to a dramatic further
increase of the critical coupling $\frac{U_C}{W}$ from 4 to 14. More
detailed studies \cite{THerN97}, however, show that the influence of
$F_{\vec{k}-\sigma}$ on magnetic stability strongly depends on the
lattice type. Increasing coordination number $({\rm sc} \longrightarrow 
{\rm bcc} \longrightarrow {\rm fcc})$ drastically 
diminishes the importance of the
non-locality leaving the local spin-dependent band shift $B_{-\sigma}$ as
the pushing mechanism for ferromagnetic stability. 

It can be shown \cite{HN97} that the SDA gives a qualitatively convincing
picture of bandferromagnetism . The main message is that the
spin-dependent ``band-shift'' $B_{-\sigma}$ and the lattice structure
are the most important ingredients. However, we should not forget the
disadvantages of the method. One of them is the neglect of quasiparticle
damping, which is assumed to destabilize the collective spin order.
\subsection{Quasiparticle Damping}

More or less by construction the SDA selfenergy (35) is real except for
a single $\delta$-peak in $Im \Sigma_{\vec{k}\sigma}(E+\I 0^+)$ at the
pole $E=T_0+B_{\vec{k}-\sigma}+U(1-\langle n_{-\sigma}\rangle)-\mu$. It
has, however, no influence since it falls into the Hubbard gap. The
quasiparticles are stable. We are therefore looking for an approach that retains
the obvious advantages of the SDA but improves it by a reasonable
inclusion of quasiparticle lifetime effects.

Starting point is an alloy-analogy for the Hubbard model which traces
back to Hubbard himself \cite{Hub64}. The idea is to 
consider the propagation of
a $\sigma$ electron through the lattice with the $-\sigma$ electrons
being ``frozen'' at their lattice sites and randomly distributed over
the crystal. When the $\sigma$ electron enters a lattice site it can
encounter two different situations. It can meet a $-\sigma$ electron or
not. That can be interpreted as a hopping through a fictitious binary
alloy, the two constituents of which are characterized by atomic levels
$E_{1\sigma},E_{2\sigma}$ and by concentrations
$x_{1\sigma},x_{2\sigma}$. The ``coherent potential approximation''
(CPA) is a standard method to perform the configurational average over
the alloy \cite{VKE68}. The resulting $\sigma$ selfenergy obeys the following
implicit equation \cite{NoltingBd7},
\begin{equation}
  0=\sum\limits^{2}_{p=1}x_{p\sigma}\frac{E_{p\sigma}-
    \Sigma_{\sigma}(E)-T_0}{1-\frac{1}{\hbar}G_{\sigma}(E)(E_{p\sigma}
    -\Sigma_{\sigma}(E)-T_0)}
\end{equation}
$G_{\sigma}(E)=\frac{1}{N}\sum\limits_{\vec{k}}G_{\vec{k}\sigma}(E)$ is
the local propagator. CPA is a single-site approximation, the resulting
selfenergy therefore wave vector independent. To proceed one has to
specify the two-component alloy. The natural way (``conventional alloy
analogy'' CAA) would be to use the zero-bandwidth results of the Hubbard
model: $E_{1\sigma}=T_0 , E_{2\sigma}=T_0+U , x_{1\sigma}=1-\langle
n_{-\sigma}\rangle = 1-x_{2\sigma}$. The resulting
$\Sigma_{\sigma}^{CAA}(E)$ includes quasiparticle damping but excludes
any spontaneous ferromagnetism \cite{FE73,SD75}. Does quasiparticle 
damping really
kill any spontaneous moment order? This is a serious question because it
has been shown \cite{VV92} that for infinite lattice dimensions CPA is an exact
(!) treatment of the alloy problem. On the other hand, the CAA selfenergy
disagrees with the strong as well as weak coupling behaviour (24),
(28). This discrepancy can only be removed by the conclusion that the
zero-bandwidth limit constitutes the wrong alloy analogy \cite{HN96}.

The fictitious binary alloy is indeed by no means predetermined. We
consider $E_{1,2\sigma}$ and $x_{1,2\sigma}$ at first as free parameters
and fix them via the high-energy expansions (21) and (24), which we
insert into the CPA equation (44). Then we expand eq. (44) with respect
to powers of $\frac{1}{E}$. Comparing coefficients we can use the
following set of equations to determine an ``optimum alloy analogy'':
\begin{eqnarray}
  \sum\limits_{p=1}^{2}x_{p\sigma}=1\nonumber\\
  \sum\limits_{p=1}^{2}x_{p\sigma}(E_{p\sigma}-T_0)=U\langle n_{-\sigma}
  \rangle
  \nonumber\\\sum\limits_{p=1}^{2}x_{p\sigma}(E_{p\sigma}-T_0)^2=U^2\langle
  n_{-\sigma}\rangle\nonumber\\\sum\limits_{p=1}^{2}
  x_{p\sigma}(E_{p\sigma}-T_0)^3=U^3\langle
  n_{-\sigma}\rangle+U^2B_{-\sigma}\langle 
  n_{-\sigma}\rangle (1-\langle n_{-\sigma}\rangle)
\end{eqnarray}
Deriving from these equations $E_{p\sigma}$ and $x_{p\sigma}$
automatically guarantees the correct strong-coupling behaviour. One
finds (p=1,2) (``modified alloy analogy'' MAA).
\begin{eqnarray}
\lefteqn{E_{p\sigma}^{MAA}=}\nonumber\\
T_0 + \frac{1}{2}[U+B_{-\sigma}+(-1)^{\cal P} \sqrt{(U+B_{-\sigma})^2-4\langle n_{-\sigma}
\rangle B_{-\sigma}}]\nonumber\\
\equiv(E_{p\sigma}^{SDA}(\vec{k}))_{\varepsilon(\vec{k})=T_0}\equiv f_p(T_0,U,\langle n_{-\sigma}
\rangle,B_{-\sigma})
\end{eqnarray} 
\begin{eqnarray}
x_{1\sigma}^{MAA}=
\frac{B_{-\sigma}+T_0+U(1-\langle n_{-\sigma}
\rangle)
-E^{MAA}_{1\sigma}}{E_{2\sigma}^{MAA}-E_{1\sigma}^{MAA}}=1-x^{MAA}_{2\sigma}\\
x_{p\sigma}^{MAA}\equiv (\alpha^{SDA}_{p\sigma})_{\varepsilon(\vec{k})=T_0}
\equiv g_p(T_0,U,\langle n_{-\sigma} \rangle,B_{-\sigma})\nonumber
\end{eqnarray} 
Surprisingly, the energies and weights coincide exactly with the
corresponding SDA entities if $\varepsilon (\vec{k})$ is simply replaced
by $T_0$. Because of the single-site aspect of the CPA 
\cite{VKE68,NoltingBd7} the
``band-correction'' $B_{\vec{k}-\sigma}$ is restricted here to its local
part $B_{-\sigma}$, the decisive band shift. Inserting (46) and (47)
into (44) yields the MAA selfenergy for all E which exhibits some
remarkable features \cite{HN96}:
\begin{itemize}
\item[(1)] As a CPA result the MAA includes quasiparticle damping 
($Im\Sigma_{\sigma}^{MAA}(E)\not= 0$), and that without giving up the
advantages of the SDA.

\item[(2)] The expectation values $\langle n_{-\sigma}\rangle$ and $B_{-\sigma}$
are to be determined selfconsistently. In principle, they take care for
a carrier concentration-, temperature- and spin-dependence of the atomic
data (46),(47) of the alloy constituents. Furthermore, and that is quite
an important aspect, $B_{-\sigma}$ brings into play in a certain sense
the itineracy of the $-\sigma$ electrons (``correlated'' electron
hopping), completely neglected in the CAA.

\item[(3)] Strong-coupling and high-energy behaviour are correctly reproduced,
more or less by construction.

\item[(4)] The general CPA theory \cite{VKE68} comes for the so-called split-band
regime (here, $U \gg W$) to the conclusion that the spectral density $S_
{\vec{k}\sigma}(E)$ should consist of two separated peaks with centres
at
\begin{equation}
T_{\vec{p}\sigma}^{CPA}=E_{\vec{p}\sigma}+x_{\vec{p}\sigma}(\varepsilon
(\vec{k})-T_0); \qquad  p=1,2
\end{equation}
Inserting (46) and (47) for $U \gg W$ yields exactly the Harris-Lange
results (6-8), a further strong confirmation of the MAA. 

\item[(5)] Contrary to the CAA the MAA allows for spontaneous
bandferromagnetism.
\end{itemize}
For a typical example the possibilities of the MAA are demonstrated in
Fig.3, which shows $S_{\vec{k}\sigma}(E)$ for strongly correlated
electrons $(\frac{U}{W}=5)$ on an fcc lattice. For less than half-filled
bands ($n<1$) the system is paramagnetic. The band occupation n=1.6 used
in Fig.3 allows for a spontaneous collective order provided U exceeds a
critcal value. Two types of splitting occur. At first the spectral
density consists for each $\vec{k}$-vector of a high-energy and a
low-energy peak separated by an energy of order U. The finite widths of
the peaks are due to quasiparticle damping and obviously energy-,
wave-vector-, spin- and temperature-dependent. The spectral weight
(area) of the low-energy peak scales with the probability that the
propagating ($\vec{k},\sigma$)-electron in the more than half-filled
band enters an empty lattice site, the weight of the upper peak that it
meets anywhere a $-\sigma$ electron. This splitting is general and not at
all bound to ferromagnetism. It demonstrates the correct strong-coupling
behaviour (6,7). Ferromagnetism appears when the two spectral density
peaks show a spin asymmetry. For the low temperature example $T=100K$
(Fig.3a) the system is close to saturation ($m=2-n$), i.e. the up-spin
states are almost fully occupied. A down-spin electron cannot avoid to
meet an up-spin electron at every lattice site being therefore forced to
perform a Coulomb interaction. Consequently the low-energy peak of 
$S_{\vec{k}\downarrow}(E)$ disappears. At higher temperatures (Fig.3b)
the peak reappears because of a partial demagnetisation, i.e. a finite
density of holes in the up-spin spectrum. At low temperatures the
high-energy peaks of $S_{\vec{k}\downarrow}(E)$ are very sharp indicating
long-living quasiparticles. A $\downarrow$-hole has no chance to meet an
$\uparrow$-hole to be scattered.
\begin{figure}[t]
  \begin{center}
    \epsfig{file=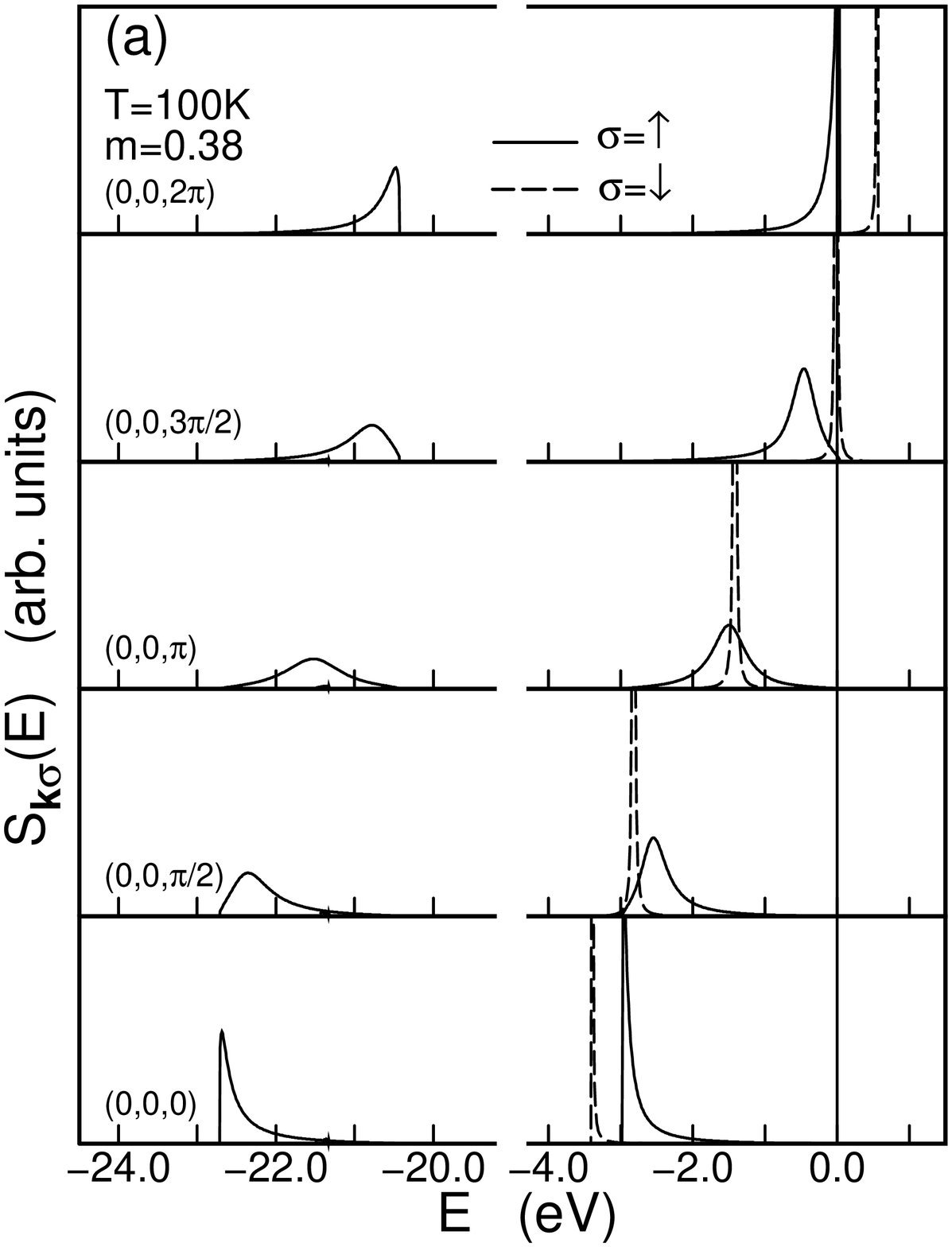,scale=0.25}
    \epsfig{file=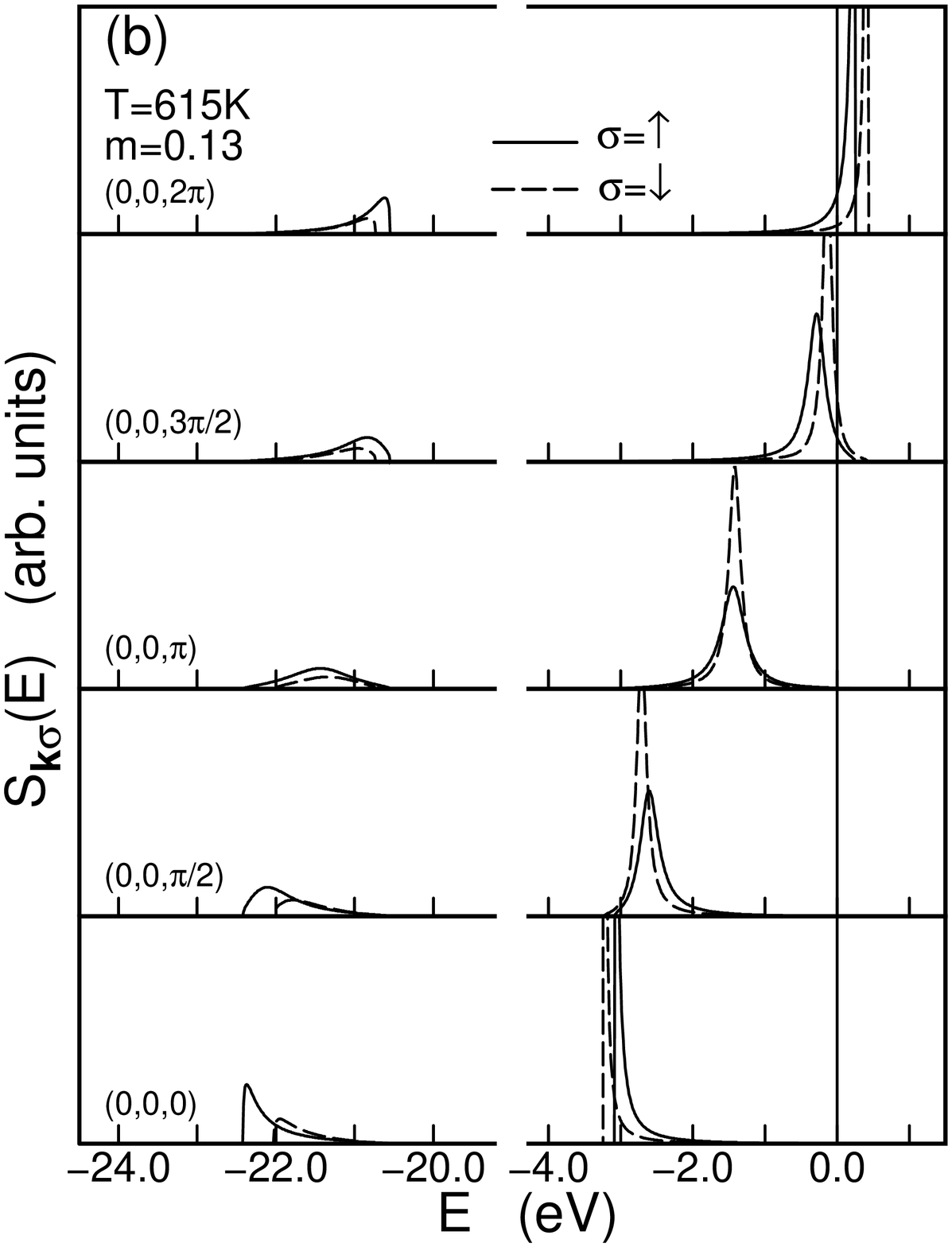,scale=0.25}
    \caption{Spectral density as a function of energy for an fcc lattice
      calculated within the MAA, (a) $T=100K$, (b) $T=615K$ for different $\vec{k}$-vectors equidistant along the (001)-direction of the 1. Brillouin zone. Further parameters: $n=1.6,\quad U=20eV,\quad W=4eV$. The vertical line indicates the position of the chemical potential.}
  \end{center}\label{fig:Nol:specdens}
\end{figure}

An interesting $\vec{k}$-dependence of the exchange splitting is
observed. At the branch-top (X-point) a ``normal'' splitting apppears,
i.e. the down-spin peak is above the up-spin peak. At the bottom
($\Gamma$-point), however, the $\uparrow$-peak is higher in energy than
the $\downarrow$-peak (``inverse exchange splitting''). The
quasiparticle dispersions of the two spin-parts are crossing as a function
of $\vec{k}$. This behaviour is a result of two competing correlation effects. 
The one is the spin-dependent exchange shift of the centres of gravity of the 
quasiparticle subbands, the other a spin-dependent band narrowing. The 
latter may overcompensate the first. In any case the exchange splitting 
exhibits a strong wave-vector dependence, even with a sign change. 

We can conclude that the MAA improves the SDA in a systematic manner by 
including quasiparticle damping. As will be discussed in Sect. 4, the main 
modification consists in a substantial destabilisation of ferromagnetism in 
the Hubbard model. One reason is the finite overlap of the spectral density 
peaks, in the SDA because of (34) excluded, which weakens the ferromagnetic 
solution of the selfconsistently evaluated model theory. 

The second shortcoming of the SDA, the incorrect weak-coupling behaviour, 
persists in the MAA. In the following we have to inspect its consequences 
for the strong-coupling phenomenon ferromagnetism. 

\subsection{Weak-coupling behaviour}
Neither the SDA selfenergy (35) nor the MAA selfenergy does fulfill the 
weak-coupling  expansion (28). We are now looking for an approach which 
interpolates reasonably between the strong- and weak-coupling regimes. 
It should extend the preceding approaches SDA and MAA by inclusion of weak 
coupling and Fermi liquid behaviour without giving up the convincing 
essentials from the SDA and MAA: spin-dependent band shift $B_{-\sigma}$ 
and quasiparticle damping. Following an idea of Kajueter and Kotliar 
\cite{KK96} 
we start with a selfenergy ansatz \cite{PWN97}, which we call the ``modified 
pertubation theory'' (MPT): 
\begin{equation}
  \Sigma_{\vec{k}\sigma}(E)=U\langle n_{-\sigma}\rangle + 
  \frac{a_{\vec{k}\sigma}U^2 \Delta_{\vec{k}\sigma}(E)}
  {1-b_{\vec{k}\sigma}U^2 \Delta_{\vec{k}\sigma}(E)}
\end{equation}
The second order contribution $\Delta_{\vec{k}\sigma}(E)$ is defined in eq. 
(29). We use in $\Delta_{\vec{k}\sigma}(E)$ for the spectral densities the 
Hartree-Fock version (30) with the selfconsistency condition (31). By 
construction the selfenergy (49) is correct in the weak-coupling regime up to 
order $U^2$. In order to fulfill simultaneously the strong-coupling regime we 
expand (49) in powers of the inverse energy $1/E$ and compare this with the 
exact selfenergy expansion (24). The first two terms are automatically 
fulfilled, the third and the fourth fit the coefficients 
$a_{\vec{k}\sigma},\;b_{\vec{k}\sigma}$ \cite{PWN97}. For the results 
presented in the 
next section we have implemented the MPT (49) in a ``dynamical mean field 
theory'' (DMFT) procedure \cite{PWN97,WPN98}. The main assumption is 
then a local 
selfenergy $\Sigma_{\vec{k}\sigma}(E)\to \Sigma_{\sigma}(E)$ , exact in 
infinite lattice dimensions \cite{MV89}. Furthermore, 
we exploit the fact that the 
Hubbard problem can be mapped under such conditions on the single-impurity 
Anderson model (SIAM) (see contribution of D. Vollhardt ) as long as a special 
selfconsistency condition is fulfilled \cite{PWN97}. 
The idea of the MPT is then applied to the simpler SIAM problem. It can
be shown, however, that the direct application of the MPT-concept to the
Hubbard model gives almost the same results.
\begin{figure}[t]
  \begin{center}
    \epsfig{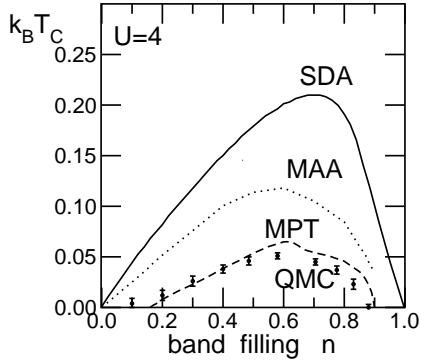}
    \caption{Curie temperature as a function of band ``filling'' (hole
      concentration) for an fcc-type $d=\infty$-lattice (51). The points
      with error bars are the Quantum-Monte-Carlo results from ref. [3]. SDA, MAA and MPT are explained in the text.}
    \label{fig:Nol:tcofn}
  \end{center}
\end{figure}

The MPT fulfills a maximum number of limiting cases: The weak-coupling 
behaviour is correct up to $U^2$-terms, for all bandoccupations $n$ 
Fermi liquid properties are recovered. The Luttinger theorem, 
\begin{equation}
  S_{ii\sigma}(0)\stackrel{!}{=} S^{(0)}_{ii\sigma}(0)\;\leftrightarrow \; 
  \mu = \mu|_{U=0}+\Sigma_{\sigma}(0)\;,
\end{equation}
is fulfilled for all bandoccupations $n$, at least if not too far from 
half-filling ($n=1$). For low temperatures a Kondo resonance appears at the 
chemical potential $\mu$. The zero-bandwidth limit ($W\to 0$) comes out 
exact for all $n$ as well as the strong-coupling behaviour gathered in 
eqs. (6)-(8). Eventually bandferromagnetism is possible within the framework 
of MPT. According to the number of correctly reproduced limiting cases the 
MPT appears to be an optimum approach to the many-body problem of the Hubbard 
model. 

\section{Discussion}
Let us compare the analytical approaches presented in the preceding Section 
with respect to the most important magnetic key-quantity, the Curie 
temperature $T_{\rm C}$. For comparison with the numerically essentially exact 
Quantum Monte Carlo results of Ulmke \cite{Ul98} we have evaluated the three theories 
for an fcc-$d=\infty$ type lattice described by the Bloch density of states: 
\begin{equation}
  \rho_0(E)= \frac{\exp{\big( -\frac{1}{2}(1-\frac{\sqrt{2}E}{t^{\ast}})
      \big)}}{t^{\ast}\sqrt{\pi (1-\frac{\sqrt{2}E}{t^{\ast}})}}
\end{equation}
The energy unit is chosen to $t^{\ast}=t\sqrt{2d(d-1)}\stackrel{!}{=}1$. 
The density of states (51) is strongly asymetric with a square root divergency 
at the upper edge. 

All the presented methods predict that ferromagnetism exists only for more 
than half-filled energy bands. In Fig. 4 bandfilling therefore means the 
hole-density and the DOS is that for holes following from (51) by 
$t^{\ast} \to -t^{\ast}$.
\begin{figure}[tp]
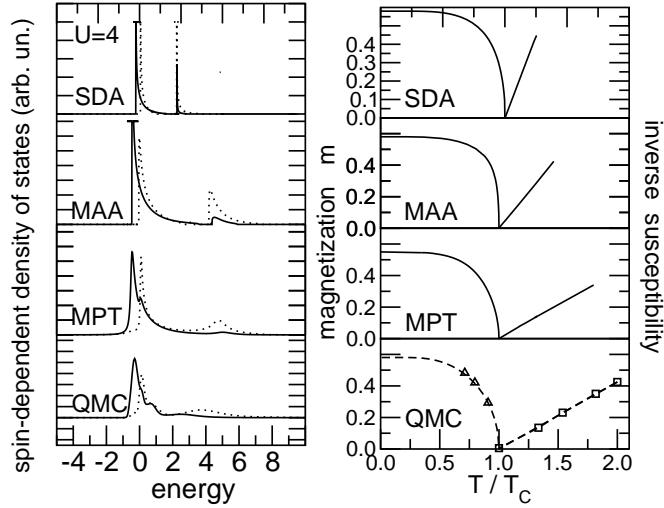

  \begin{center}
    \epsfig{file=fedos_1.eps,scale=0.35}
    \epsfig{file=mandchi.eps,scale=0.35}
    \caption{Left: Quasiparticle density of states as function of energy
      for the fcc-type lattice (51). Solid lines for $\sigma
      =\uparrow$, dotted lines for $\sigma =\downarrow$. QMC results
      from [3]. Temperature T is chosen so that for all presented
      methods a magnetization $\rm{m}=0.4$ is found. Right: For the
      same parameters as in the left part the magnetisation m and the
      inverse paramagnetic susceptibility $\chi^{-1}$ as function of the
      reduced temperature $T/T_{\rm{C}}$. Bottom figure: QMC results,
      dashed line for $T/T_{\rm{C}}\le 1 : S=1/2$-Brillouin function.}
    \label{fig:Nol:dos-chi}
  \end{center}
\end{figure}

The role of the spin-dependent band shift $B_{-\sigma}$ (10) for the magnetic 
stability becomes evident by comparison of the three methods with their 
$B_{-\sigma}=0$ counterparts. That is for SDA the Hubbard I-solution 
\cite{Hub63} which 
yields ferromagnetism only for very asymmetric DOS. In the case of (51) there 
appears ferromagnetism in a strongly restricted region of very low hole 
densities. The band shift $B_{-\sigma}$ in the SDA obviously leads to a 
drastic enhancement of ferromagnetic stability. The counterpart of MAA 
is the conventional alloy analogy (CAA) that does not allow for any $n$ 
ferromagnetic order \cite{FE73,SD75}. Finally, the MPT-counterpart, that neglects the 
bandshift, is the Kajueter-Kotliar approach \cite{KK96}. It indeed exhibits 
ferromagnetism but with substantially lower Curie temperatures $T_{\rm C}$.

The influence of quasiparticle damping can best be judged by comparing the 
results of SDA and MAA. The inclusion of lifetime effects in MAA presses 
the Curie temperature to half the SDA-values. The ferromagnetic coupling 
strength is therefore substantially weakened by quasiparticle damping. 
The incorporation of the correct weak-coupling behaviour by MPT leads to 
a further $T_{\rm C}$-reduction possibly because of the screening
tendency with respect to the effective magnetic moments manifesting itself in the appearance of a Kondo resonance. The MPT-results for $T_{\rm C}(n)$ are 
closest to the essentially exact numerical results of Ulmke \cite{Ul98}, which 
have been read off from the zeros of the inverse paramagnetic susceptibility. 

The similarities and the differences of the various methods come out by the 
quasiparticle density of states in Fig. 5a. All theories show the splitting 
into the two Hubbard bands and the additional exchange splitting which 
takes care for the spin asymmetry. As a consequence of the neglect of damping 
effects the DOS-structures are rather sharp in the SDA. They appear smoother 
in MAA. The correct low-energy behaviour of the MPT leads to the best approach 
to the Q-DOS found by ``Quantum Monte Carlo'' (QMC)-evaluation of a 
``Dynamical Mean Field Theory''. Note that for a reasonable comparison of 
the Q-DOS we have chosen in Fig. 5a for the various theories temperatures 
which lead to the same magnetization $m=0.4$. 

Fig. 5b demonstrates impressively the qualitative equivalence of the 
presented methods when we plot the spontaneous magnetization and the 
paramagnetic susceptibility as function of the reduced temperature 
$T/T_{\rm C}$. There are indeed strong similarities. All magnetization curves are 
Brillouin function-like with full polarization at $T=0$. The only exception 
is MPT, which shows up a slight deviation from saturation at $T=0$. For the 
chosen parameter set all methods predict a second order transition at 
$T_{\rm C}$. The paramagnetic susceptibility follows in all theories a ideal 
Curie-Wei{\ss} behaviour: $\chi=C (T-\Theta)^{-1}$. The paramagnetic 
Curie temperature $\Theta$ equals in any case the Curie temperature 
$T_{\rm C}$ and even the Curie constants $C$ are all very similar: 
$C=$0.42 (SDA), 0.52 (MAA), 0.57 (MPT), 0.47 (QMC). One important 
consequence of this striking qualitative equivalence and correctness of SDA, 
MAA and MPT compared to QMC is that even the rather simple SDA can be used 
to describe the magnetism of more complicated structures (films, surfaces, 
multilayer, real substances). 

We conclude that ferromagnetism does exist in the Hubbard model depending on 
lattice structure, band occupation, Coulomb coupling and temperature. By 
a critical comparison of three different analytical approaches we could 
demonstrate the essentials for ferromagnetic stability, in a positive sense 
the spin-dependent band shift $B_{-\sigma}$, which represents the correlated 
hopping of opposite spin electrons, and in a negative sense the finite 
lifetime of quasiparticles, which results in a drastic lowering of the 
ferromagnetic coupling strength. The correct inclusion of Fermi liquid 
properties eventually leads to the most convincing description of 
ferromagnetism in the Hubbard model.
\end{document}